\documentclass[aps,prc,reprint,groupedaddress]{revtex4-1}

\usepackage[dvipdfmx]{graphicx}
\newcommand{\disregard}[1]{} 

\begin{document}


\title{
Analytical formula for
numerical evaluations of the Wigner rotation matrices at high spins
}

\author{Naoki Tajima}
\email[]{tajima@quantum.apphy.u-fukui.ac.jp}
\affiliation{Department of Applied Physics, University of Fukui, Bunkyo 3-9-1, Fukui, 910-8507, Japan}

\date{\today}

\begin{abstract}
The Wigner $d$ function, which is the essential part of an irreducible representation
of SU(2) and SO(3) parameterized with Euler angles, has been know to suffer from a serious 
numerical errors at high spins, if it is calculated by means of the Wigner formula
as a polynomial of cos and sin of half of the second Euler angle.
This paper shows a way to avoid this problem by expressing the $d$ functions
as the Fourier series of the half angle.
A precise numerical table of the coefficients of the series is provided
as Supplemental Material.
\end{abstract}


\maketitle

\section{Introduction \label{intro}}

The matrix elements of the rotation operator
between angular-momentum eigenstates are called the Wigner $D$ function \cite{MERose}.
When the rotation is specified with the three Euler angles
($\phi$, $\theta$, $\psi$), 
and the eigenstates are labeled with the magnitude
($j$ = $0, \frac{1}{2}, 1, \frac{3}{2}, \cdots$) 
and the $z$ component 
($m$ or $k$ = $-j, -j+1, \cdots, j$)
of the angular momentum vector, 
the $D$ function can be decomposed into three factors,
\begin{eqnarray}
D^{j}_{mk}(\phi,\theta,\psi) 
& = &
\langle j m \vert e^{-i \phi \hat{j}_z} e^{-i \theta \hat{j}_y} e^{-i \psi \hat{j}_z} 
\vert j k \rangle \nonumber \\
& = & 
e^{-i (m \phi + k \psi)} d^{j}_{mk}(\theta),  \label{eq:Dfunc}
\end{eqnarray}
where
\begin{equation} \label{eq:dfunc}
d^{j}_{mk}(\theta) =
\langle j m \vert e^{-i \theta \hat{j}_y} \vert j k \rangle
\end{equation}
is the nontrivial part which needs some method to evaluate.
It is called the Wigner (small) $d$ function.
In the standard phase convention for angular-momentum eigenstates,
the matrix elements of $\hat{j}_y$ are purely imaginary and thus
$d^{j}_{mk}(\theta)$ takes on real numbers.

One may be anxious about the fact that Bohr and Mottelson\cite{BM69} define the rotation matrix
$D^{j}_{mk}$ as the complex conjugate of the right-hand side (r.h.s.) of Eq.(\ref{eq:Dfunc}).
However, their definition for $d^{j}_{mk}$ is identical with Wigner's and the
definition of the $d$ function is unique.

The Wigner $D$ function is used in various fields of physics.
In some applications,
those for large values of $j$ (say, $>50$) are necessary.
An example in nuclear structure physics is
the projection from spatially deformed solutions of modern 
realistic mean-field models
to eigenstates of large angular momentum.
For toy models, too, 
one occasionally needs $d$ functions for very large $j$ 
to confirm the validity of one's picture
under extreme conditions.
(e.g., a two-rotor model of Refs.~\cite{Taj13} and \cite{TO11}).

The explicit form of the $d$ function is given by the Wigner formula\cite{MERose},
which can be written as,
\begin{equation} \label{eq:Wigner_formula}
d^{j}_{mk}(\theta) = \sum_{n=n_{\rm min}}^{n_{\rm max}} (-1)^{n} W_{n}^{jmk}(\theta),
\end{equation}
where
$n_{\rm min}$ and $n_{\rm max}$ are zero or positive integers,
\begin{eqnarray}
n_{\rm min} & = & \max (0, k-m), \label{eq:nmin} \\
n_{\rm max} & = & \min (j-m,j+k), \label{eq:nmax}
\end{eqnarray}
and
\begin{equation}
W_{n}^{jmk}(\theta)
=
w_{n}^{jmk} \cdot
\left( \cos {\textstyle \frac{\theta}{2}} \right)^{2j+k-m-2n}
\left(-\sin {\textstyle \frac{\theta}{2}} \right)^{m-k+2n}
\label{eq:Wigner_formula_term},
\end{equation}
\begin{equation} \label{eq:small_w_jmkn}
w_{n}^{jmk} = \frac{\sqrt{(j+m)!(j-m)!(j+k)!(j-k)!}} {(j-m-n)!(j+k-n)!(n+m-k)!n!} .
\end{equation}
However, this formula suffers from a serious loss of precision at high spins
(i.e., for large $j$) except in the neighborhood of $\theta = 0, \pi$.

For example, assuming that $j$ is a positive integer,
$\theta = \frac{\pi}{2}$, and $m=k=0$, one obtains
\begin{equation}
W_{n}^{j00}\left(\frac{\pi}{2}\right)=\frac{1}{2^j} \left[\frac{j!}{(j-n)!n!}\right]^2,
\end{equation}
which, if $j$ is even, becomes maximum at $n=\frac{j}{2}$,
\begin{equation} \label{eq:maxwigterm}
W_{j/2}^{j00}\left(\frac{\pi}{2}\right)=
\frac{1}{2^j} \frac{j!}{\left[(j/2)!\right]^2}
\approx \sqrt{\frac{2}{\pi j}} 2^j .
\end{equation}
(The Stirling's formula is used in the last approximation.)
The absolute value of the $d$ function is not greater than one 
because it is a matrix element of a unitary operator between normalized states.
Wigner's formula expresses the $d$ function
as a result of cancellation among terms
of possibly huge size, $W_n^{jmk} \sim 2^j$.
For $j \sim 54$, the precision of double-precision floating-point numbers (53-bit mantissa) is
lost completely,
and even quadruple precision float numbers (113-bit mantissa) is lost completely for
$j\sim 114$.

A few remedies have been proposed \cite{CIG99,Dac06}
but the results are not completely free of
the precision loss.
In this paper, I investigate the details of this loss of significance, and then
present a perfect remedy to avoid such numerical difficulty.

\section{The Fourier-series expression of $d$ functions}

One can see easily that terms like
$\cos^{\lambda} \frac{\theta}{2} \sin^{\mu} \frac{\theta}{2}$
appearing in the r.h.s.\ of Eq.~(\ref{eq:Wigner_formula_term}),
where $\lambda$ and $\mu$ are zero or positive integers such that 
$\lambda + \mu \le 2j$,
can be expressed as
linear combinations of terms 
$\sin\frac{\kappa \theta}{2}$ (when $\mu$ is odd)
and $\cos\frac{\kappa \theta}{2}$ (when $\mu$ is even)
with integers $\kappa$ in $0 \le \kappa \le \lambda + \mu$ and 
$\kappa \equiv \lambda + \mu$ (mod 2),
by means of repeated applications of 
the elementary trigonometric identities called the 
product-to-sum identities or the prosthaphaeresis formulas. 

Because the power of $\sin\frac{\theta}{2}$ is $\mu = m-k+2n$
in Eq.~(\ref{eq:Wigner_formula_term}),
one can see that $\mu \equiv m-k$ (mod 2)
and that the $d$ function is an even (odd) function
if $m-k$ is even (odd), 
which can be expanded only with cos (sin) function.
This may also be deduced from Eq.~(\ref{eq:mk_km_minus_t}),
one of the properties of the $d$ function
which I have enumerated in the appendix \ref{sec:dfunctionproerties}.

From these considerations,
one can conclude that the Fourier expansion of the $d$ function has the following form:
\begin{equation} \label{eq:DfuncFourierSeries}
d^{j}_{mk} (\theta) = \sum_{\nu} t^{jmk}_{\nu} f(\nu \theta),
\end{equation}
where
\begin{equation}  \label{eq:Dfunc_f_cos_sin}
f = \left\{\begin{array}{c} \cos \\ \sin \end{array}\right\}
\;\;\; \mbox{for} \;\;
\left\{\begin{array}{c} \mbox{even} \\ \mbox{odd} \end{array}\right\}
m-k,
\end{equation}
and the summation runs over
\begin{equation} \label{eq:Dfunc_n_runs_over}
\nu=\nu_{\rm min}, \nu_{\rm min}+1, \cdots,j,
\end{equation}
with the values of $\nu_{\rm min}$ given in Table~\ref{tab:nmin}.

\begin{table}[hbt]
\caption{\label{tab:nmin}
The values of $\nu_{\rm min}$ appearing in Eq.~(\ref{eq:DfuncFourierSeries})
}
\begin{center}
\begin{tabular}{|c|c|c|}
\hline
          & Even $m-k$ & Odd $m-k$ \\ 
\hline
Even $2j$ & 0 & 1 \\
\hline
Odd  $2j$ & $\frac{1}{2}$ & $\frac{1}{2}$ \\
\hline
\end{tabular}
\end{center}
\end{table}

For example, for $j=\frac{7}{2}$ and $m=-k=\frac{1}{2}$,
the Wigner formula (\ref{eq:Wigner_formula}) gives an expression,
%
\newcommand{\halftheta}{{\textstyle \frac{\theta}{2}}}
\newcommand{\fractheta}[2]{{\textstyle \frac{#1 \theta}{#2}}}
\begin{eqnarray}
d^{\frac{7}{2}}_{\frac{1}{2},-\frac{1}{2}}(\theta) &=&
\sin ^7 \halftheta
-12 \cos^2 \halftheta\,\sin ^5 \halftheta
+18 \cos ^4 \halftheta\,\sin ^3 \halftheta
\nonumber \\
& & -4\cos ^6 \halftheta\, \sin \halftheta ,
\label{eq:dfuncexwigner}
\end{eqnarray}
which can be rewritten in the form (\ref{eq:DfuncFourierSeries}) as
\begin{equation} \label{eq:dfuncexfs}
d^{\frac{7}{2}}_{\frac{1}{2},-\frac{1}{2}}(\theta) =
-\frac{35\,\sin \fractheta{7}{2} -5\,\sin \fractheta{5}{2}
+15\,\sin \fractheta{3}{2}-9\,\sin \halftheta}{64}.
\end{equation}


By utilizing the orthogonality of $\cos \nu \theta$ and $\sin \nu \theta$ over
$0 \le \theta \le 4 \pi$ (considering that $\nu$ can take both integer and half-integer values),
one can express the coefficients $t_{\nu}^{jmk}$ by an integral
\begin{equation} \label{eq:DfuncFourierSeriesCoef}
t^{jmk}_{\nu} = \frac{1}{2 \pi (1+\delta_{\nu 0})} \int_{0}^{4\pi} d^{j}_{mk}(\theta)
f(\nu \theta) d\theta ,
\end{equation}
where
$\delta_{\nu 0}=1$ for $\nu=0$ and $\delta_{\nu 0}=0$ for $\nu=\frac{1}{2},1,\frac{3}{2},\cdots$.

By substituting the $d$ function in Eq.~(\ref{eq:DfuncFourierSeriesCoef}) with
Eqs.~(\ref{eq:Wigner_formula}-\ref{eq:small_w_jmkn}),
I have derived a more useful expression for $t_{\nu}^{jmk}$ 
containing only four elementary operations of arithmetic,
\begin{eqnarray}
t_{\nu}^{jmk} & = & 
\frac{2(-1)^{m-k}}{1+\delta_{\nu 0}} 
\sum_{n=n_{\rm min}}^{n_{\rm max}} (-1)^{n} w_{n}^{jmk}
\sum_{r=0}^{[\nu-\frac{1}{2}p]} (-1)^{r} 
\mbox{\footnotesize $
  \left(
    \begin{array}{c}
      2 \nu \\
      2r+p
    \end{array}
  \right)
$}
\nonumber \\
 & \times & \frac{1}{2\pi}I_{2(j+\nu-n-r)-m+k-p,2(n+r)+m-k+p}
 \label{eq:tCoefAnal}
\end{eqnarray}
where 
$n_{\rm min}$ and $n_{\rm max}$ are those already defined by Eqs.~(\ref{eq:nmin}) and (\ref{eq:nmax}),
the square brackets are the floor function, i.e.,
$[l+x]=l$ for integer $l$ and real $x$ in $[0,1)$,
\begin{equation}
p \equiv \left\vert m-k \right\vert \;\;\; ({\rm mod} \; 2),
\end{equation}
i.e., 
\begin{eqnarray}
p = 0 & \;\;\; \mbox{for} \;\;\; & k=m, m \pm 2, m \pm 4,\cdots,\\
p = 1 & \;\;\; \mbox{for} \;\;\; & k=m \pm 1, m \pm 3,\cdots,
\end{eqnarray}
and
\begin{equation} \label{eq:integral1}
I_{\lambda \mu} = \int_{0}^{2\pi} \cos^{\lambda}x \sin^{\mu} x \,dx
\end{equation}
with zero or positive integers for $\lambda$ and $\mu$. 
If both $\lambda$ and $\mu$ are even,
\begin{equation} \label{eq:integralFormula1}
I_{\lambda \mu} = \frac{2\pi(\lambda-1)!! (\mu-1)!!}{(\lambda+\mu)!!},
\end{equation}
while $I_{\lambda \mu}=0$ otherwise.

Unlike the r.h.s.\ of Eq.~(\ref{eq:Wigner_formula}),
where the terms can have huge sizes and thus the numerical error is a serious problem,
the r.h.s.\ of Eq.~(\ref{eq:DfuncFourierSeries}) is a summation of
terms of order one or less and hence the problem is expected to disappear.
This can be seen by calculating the integrals of the squares of the both sides of
Eq.~(\ref{eq:DfuncFourierSeries}),
\begin{eqnarray} 
\int_{0}^{4\pi} d^{j}_{mk}(\theta)^2 d\theta & = &
\sum_{\nu} \sum_{\mu} t_{\nu}^{jmk} t_{\mu}^{jmk} 
\int_{0}^{4\pi} f(\nu \theta) f(\mu \theta) d\theta
\nonumber \\
& = &
\sum_{\nu} \frac{4\pi}{2-\delta_{\nu 0}}
\left(t_{\nu}^{jmk}\right)^2.
\label{eq:perceval}
\end{eqnarray}
Because the absolute values of $d$ functions are $\le 1$,
the left-hand side is $\le 4\pi$ and, consequently, it holds
$|t_{\nu}^{jmk}| \le 2-\delta_{\nu 0}$.

A further study from the numerical point of view has indicated that
the maximum (among all the possible combinations of $m$, $k$, and $\nu$)
value of $|t_{\nu}^{jmk}|$ is $1$ for $j \le 1$,
decreases as $j$ increases from an integer to the next half integer,
and does not change as $j$ increases from a half integer to the next integer.
For the interval $50 \le j \le 100$, the maximum value for integer $j$ behaves
as $\approx 1.13/\sqrt{j}$.

\section{Computation of the numerical values of the coefficients}

Unfortunately, Eq.~(\ref{eq:tCoefAnal}) also suffers from a serious
loss of significant digits in ordinary floating-point numerical calculations.
Indeed, for even integer $j$,
the term having the maximum magnitude
among those in the r.h.s.\ of Eq.~(\ref{eq:tCoefAnal}) occurs 
at $m=k=0$, $n=r=\frac{1}{2}j$, $\nu=j$,
with the maximum value of
\begin{equation} \label{eq:maxfsterm}
\frac{1}{\pi} w^{j00}_{j/2} 
\mbox{\footnotesize $\left(\begin{array}{c} 2j \\ j  \end{array}\right)$}
I_{2j,2j}
\approx
\frac{2^{2(j+1)}}{(\pi j)^2},
\end{equation}
which is roughly $2^j$ times as large as
the value given by Eq.~(\ref{eq:maxwigterm}).

To avoid this problem,
I first evaluated the r.h.s.\ of Eq.~(\ref{eq:tCoefAnal}) 
rigorously as rational numbers or square root
of rational numbers by means of a formula-manipulation software MAXIMA.
Numerical values to be used in programs coded in Fortran, C, etc.,
can be calculated from such rigorous numbers
to the full precision of the 64-bit floating-point number.

However, the computation time turned out to be excessively long
for large values of $j$.
To speed up the computation,
I have changed the method to evaluate Eq.~(\ref{eq:tCoefAnal}) not
rigorously but in terms of high-precision floating-point numbers (of
MAXIMA).  
This does not seem to be a major drawback
because numerical values are sufficient
for most of practical purposes.

I change the precision of floating-point numbers 
depending on $j$ in such a way that
the number of digits equals the common logarithm of
the value of Eq.~(\ref{eq:maxfsterm}) divided by $10^{-18}$.
It increases with $j$, reaching 74 digits for $j=100$.  
Precise 64-bit floating-point numbers can be obtained simply
by truncating the high-precision results.

Empirically, the time necessary to compute all the
coefficients $t^{jmk}_{\nu}$ for each $j$ increases as $j^4$.  
The new method takes
43 h for $j=100$ with a personal computer with a CPU Intel
core-i7 3960X running at 3.3GHz, using one physical core.

I use the obtained 64-bit floating-point number coefficients to evaluate
the Fourier-series formula for $d$ functions.  I provide data files
of the numerical values of the coefficients, together with a sample
FORTRAN90 program to read the data and calculate the values of the $d$
function, as Supplemental Material \cite{SuppMat}.

For each value of $j$,
there are $(2j+1)^2$ possible combinations of the values of
$m$ and $k$ (because $-j \le m \le j$, $-j \le k \le j$).
Only about a quarter of them are independent, however,
because of the properties of the $d$ function expressed by
Eqs.~(\ref{eq:mk_minus_km}), (\ref{eq:mk_km}), and (\ref{eq:mk_minus_mk}) 
in the appendix \ref{sec:dfunctionproerties}. 
Therefore,
I consider only such combinations as
$m \ge 0$ and $k \le |m|$ in the following analysis of the numerical precision.

In other words, the coefficients $t_{\nu}^{jmk}$ have the following symmetry,
\begin{eqnarray} 
t_{\nu}^{jkm}     & = & (-1)^{m-k} t_{\nu}^{jmk}, \label{eq:tcoefreduction1} \\
t_{\nu}^{j,-m,-k} & = & (-1)^{m-k} t_{\nu}^{jmk}, \label{eq:tcoefreduction2} \\
t_{\nu}^{j,-k,-m} & = &            t_{\nu}^{jmk}. \label{eq:tcoefreduction3}
\end{eqnarray}
Hence, the numerical data  for $t_{\nu}^{jmk}$ in the Supplemental Material are given
only for $m \ge 0$ and $k \le |m|$.

There are 
$\approx \frac{1}{2} (j_{\rm max}+\frac{3}{2})^4$ 
coefficients for $j=0,\frac{1}{2},1,\cdots,j_{\rm max}$.
The size of the memory to store them as 64-bit floating-point numbers
amounts to 27 (404) MiB for $j_{\rm max}$=50 (100).
My data are given as text files for the sake of compatibility, whose sizes
are not very different from the above memory sizes 
after they are compressed (with the software GZIP).


The magnitude of the coefficient $t^{jmk}_{\nu}$ becomes smaller for larger values of
$j$.
However, even at $j=100$, 80\% (90\%) of the coefficients are larger than
$10^{-5}$  ($10^{-10}$).

Some of the coefficients vanish according to rules unmentioned so far. For example,
$t^{jmk}_{\nu}=0$ if $j$ is an integer, $m=0$ and/or $k=0$,
and $j-\nu \equiv 1$ (mod 2).
One can prove this vanishment by using Eq.~(\ref{eq:pi_plus_theta}).
I do not use these additional rules but simply give zero values in the
data files.

The evaluations of $\cos\,\nu\theta$ and $\sin \, \nu\theta$ should be calculated
by means of a recursion relation,
\begin{equation} \label{eq:cossinnutheta}
\left( 
  \begin{array}{c} 
    \cos\,(\nu+1)\theta \\
    \sin\,(\nu+1)\theta
  \end{array} 
\right)
=
\left( 
  \begin{array}{cc} 
    \cos\,\theta, & -\sin\,\theta \\
    \sin\,\theta, &  \cos\,\theta 
  \end{array} 
\right)
\left( 
  \begin{array}{c} 
    \cos\,\nu\theta \\
    \sin\,\nu\theta
  \end{array} 
\right),
\end{equation}
which is nothing but the trigonometric (angle) addition theorem.
For integer $j$, the initial values are $\cos\,0=1$ and $\sin\,0=0$.
For half integer $j$, one has to calculate, first, 
the initial values $\cos\,\frac{\theta}{2}$ and $\sin\,\frac{\theta}{2}$
and, second, $\cos\,\theta$ and $\sin\,\theta$ using identities
$\cos\,\theta=\cos^{2}\frac{\theta}{2}-\sin^{2}\frac{\theta}{2}$
and $\sin\,\theta=2\sin\,\frac{\theta}{2}\,\cos\,\frac{\theta}{2}$.

\begin{figure}[tbh]
\includegraphics{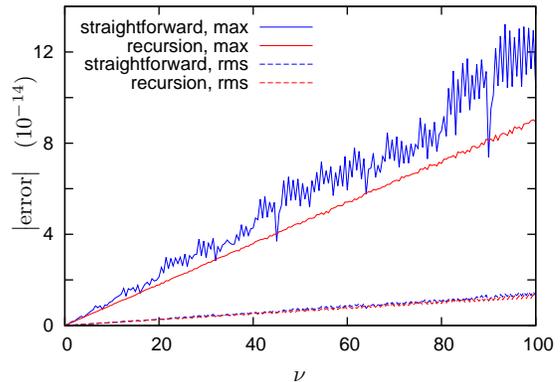}%
\caption{\label{fig:sin_cos_n_theta}
Maximum and root-mean-square errors in the 
values of  $\cos \nu \theta$ and $\sin \nu \theta$,
numerically calculated with 64-bit floating-point numbers,
over $0 \le \theta \le \pi$.
The values of $\theta$ are sampled with a uniform spacing of $10^{-5}$ degree.
The abscissa is $\nu =0, \frac{1}{2}, 1 , \cdots, 100$.
Recursion means the usage of Eq.~(\ref{eq:cossinnutheta})
together with the initial values described in the following sentences.
}
\end{figure}

Straightforward evaluation of $\cos\,\nu\theta$ and $\sin\,\nu\theta$ 
(i.e., passing the value of $\nu \theta$ to the internal functions cos and sin)
requires about $j$ calls to the functions and is computationally
very inefficient. Moreover, 
as shown in Fig.~\ref{fig:sin_cos_n_theta}, such straightforward evaluation
causes slightly larger numerical errors probably due to the loss of significant
digits in reducing the value of $\nu\theta$ to an interval such as $[-\frac{\pi}{4},\frac{\pi}{4}]$,
especially when the value of $|\nu \theta|$ is large.
The reason why the recursion formula (\ref{eq:cossinnutheta})
does not suffer from large errors even after hundred steps may be attributed
to the fact that the magnitudes of cos and sin functions are always not greater than one.

\section{Precision of the numerical values of the $d$ function \label{sec:precision}}

\begin{figure}[tbh]
\includegraphics{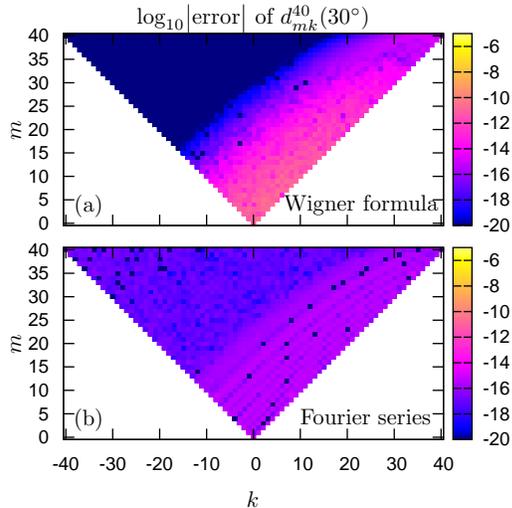}%
\caption{\label{fig:km_j40_t30}
The common logarithm of the error in the 
64-bit floating-point numerical value of the $d$ function
$d^{j}_{mk}(\theta)$ as a function of $m$ and $k$
for $j=40$ and $\theta=30^\circ$.
The value of the $d$ function is calculated by means of the Wigner formula
in part (a) and the Fourier series expression in part (b).
Only a quarter of the possible combinations of $m$ and $k$ are
plotted because the error is symmetric in the lines $m=\pm k$.
Errors smaller than $10^{-20}$ are painted with the same color
as that for the error of $10^{-20}$.
}
\end{figure}

In this section, I compare the errors of
the values of $d^{j}_{mk}(\theta)$
calculated with 64-bit floating-point numbers
according to
the Wigner formula (\ref{eq:Wigner_formula})
and the Fourier series expression (\ref{eq:DfuncFourierSeries}).
The errors have been calculated as the differences from the exact
values calculated by applying a formula-manipulation software MAXIMA
to the Wigner formula.

\begin{figure}[tbh]
\includegraphics{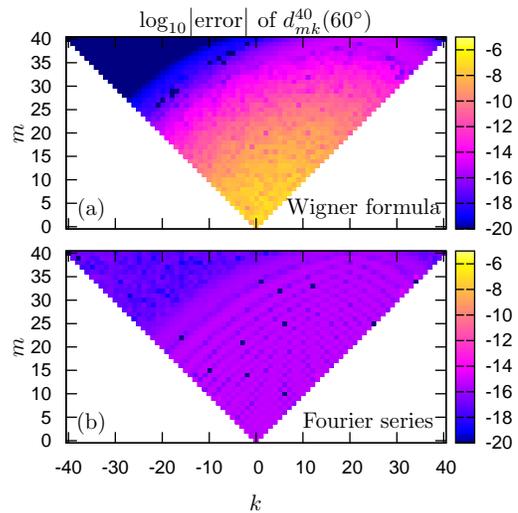}%
\caption{\label{fig:km_j40_t60}
The same as in Fig.~\ref{fig:km_j40_t30} but for $\theta=60^\circ$.
}
\end{figure}

\begin{figure}[tbh]
\includegraphics{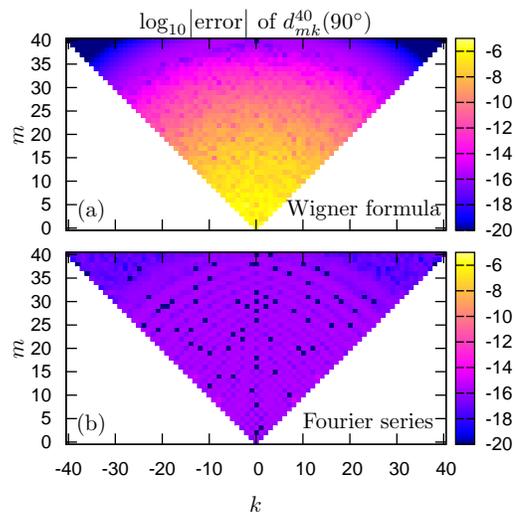}%
\caption{\label{fig:km_j40_t90}
The same as in Fig.~\ref{fig:km_j40_t30} but for $\theta=90^\circ$.
}
\end{figure}

Figures~\ref{fig:km_j40_t30}, ~\ref{fig:km_j40_t60}, and \ref{fig:km_j40_t90}
show the errors for $\theta=30^{\circ}$, $60^{\circ}$, and $90^{\circ}$, 
respectively.
The errors are expressed as a function of $(m,k)$ while $j$ is fixed at 40.
I have found that similar plots for $\theta > 90^\circ$ look almost
indistinguishable from those for $180^\circ-\theta$ except that the
sign of $k$ is reversed,
as could be foreseen from Eq.~(\ref{eq:pi_minus_theta}).

One can see that the Wigner formula results in very large errors,
up to $\sim 10^{-5}$ for $\theta \sim 90^\circ$ and $m \sim k \sim 0$,
while the Fourier series expression gives precisely 15 digits
irrespective of the values of $\theta$, $j$, $m$, and $k$.
I confidently recommend
the Fourier series expression over the Winger formula already at $j \sim 40$.  

For a special purpose, however, the Wigner formula still has an advantage.
For regions of the arguments
$\theta \sim 0^\circ$ ($180^\circ$)  and $|m+k| \sim 0$ ($2j$),
the Wigner formula has smaller errors than $10^{-15}$,
i.e., than the level of the almost constant error of the Fourier series expression.
In such regions of the arguments,
the magnitude of $d^{j}_{mk}(\theta)$ is very small and
a small number of terms dominate in the summation of the Wigner formula, while
many terms of order 1 cancel among themselves to give the small value
in the Fourier series expression.
Therefore, if one needs such high precision at those regions, it would
be a good idea to develop a program which switches between the two formulas depending on the
values of $\theta$, $j$, $m$, and, $k$.

\begin{figure}[tbh]
\includegraphics{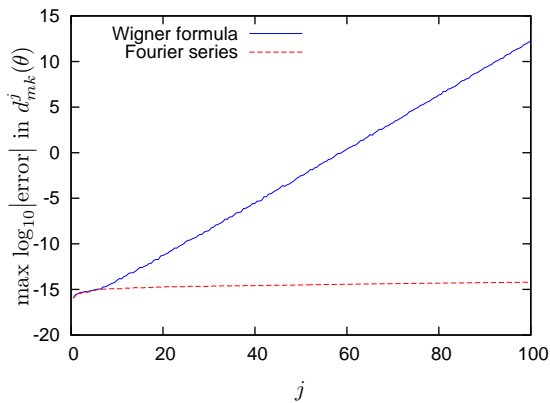}%
\caption{\label{fig:maxerr_vs_j}
The common logarithm of the maximum error in the numerical value of
the $d$ function $d^{j}_{mk}(\theta)$ versus $j$ calculated with
the Wigner formula (blue solid line) and the Fourier series
expression (red dashed line). The maximum is taken over
the values of $m$, $k$, and $\theta$ for each $j$.
}
\end{figure}

In Fig.~\ref{fig:maxerr_vs_j},
I compare the maximum errors of $d^{j}_{mk}(\theta)$
as functions of $j$.
The maximum is taken over the values of $\theta$ 
from $0^{\circ}$ to $180^{\circ}$ with an increment of $5^{\circ}$
and all possible combinations of $m$ and $k$.
For both formulas,
the maximum error increases as exponential functions of $j$.
The Wigner formula increases its error, however, far faster
than the Fourier series expression.
For the interval $20 \le j \le 100$,
the error of the Wigner formula can be approximated
by $\log_{10}|{\rm error}| \approx 0.294 j -17.2$
while that of the Fourier series expression
can be approximated
by $\log_{10}|{\rm error}| \approx 0.006 j -14.8$.
In other words,
by increasing $j$ by one, the error of the Wigner formula
is doubled, while that of
the Fourier series expression increases only by 1.4\%.

\section{Summary \label{sec:summary}}

I have shown that the Wigner formula for the $d$ function results in
intolerable large numerical errors for large values of the angular
momentum quantum number $j$.  
On the other hand, the Fourier series expression for the
$d$ function is shown to be free of such errors, providing precision
of $\sim 10^{-14}$ even at $j=100$.  An analytic expression for the
coefficients of the Fourier series is given.  Their numerical values,
which are precise as far as 64-bit floating-point numbers can express, are
provided as electric files in Supplemental Material.
Sample programs in FORTRAN90 to use the data files are also provided.






\appendix
\section{Properties of the $d$ function utilized in this paper \label{sec:dfunctionproerties}}

I enumerate the symmetries of the $d$ function to be referred to
in this paper.

First,
the unitarity of rotations (i.e, the Hermite conjugate operator is the inverse operator)
means
\begin{equation} \label{eq:unitarity}
d^{j}_{mk}(\theta)=d^{j}_{km}(-\theta).
\end{equation}
Second, 
the composition of two rotations can be rewritten as multiplication of matrices to represent them,
\begin{equation} \label{eq:composition}
d^{j}_{mk}(\theta_1+\theta_2) = \sum_{\nu=-j}^{j} d^{j}_{m\nu}(\theta_1) d^{j}_{\nu k}(\theta_2).
\end{equation}
I need three more relations,
whose easiest derivation may be to use the Wigner formula (\ref{eq:Wigner_formula}) 
as in Ref.~\cite{MERose},
\begin{eqnarray} \label{eq:}
d^{j}_{mk}(\theta) & = & (-1)^{m-k} d^{j}_{mk}(-\theta), \label{eq:mk_km_minus_t}\\
d^{j}_{mk}(\theta) & = & d^{j}_{-k,-m}(\theta), \label{eq:mk_minus_km}\\
d^{j}_{mk}(\pi) & = & (-1)^{j+m} \delta_{m,-k}. \label{eq:pi_rotation}
\end{eqnarray}
From Eqs.~(\ref{eq:unitarity}) and (\ref{eq:mk_km_minus_t}), one can prove
\begin{equation} \label{eq:mk_km}
d^{j}_{mk}(\theta) = (-1)^{m-k} d^{j}_{km}(\theta),
\end{equation}
from Eqs.~(\ref{eq:mk_minus_km}) and (\ref{eq:mk_km}),
\begin{equation} \label{eq:mk_minus_mk}
d^{j}_{mk}(\theta) = (-1)^{m-k} d^{j}_{-m,-k}(\theta),
\end{equation}
and from Eqs.~(\ref{eq:composition}), (\ref{eq:pi_rotation}), and (\ref{eq:mk_km}),
\begin{equation} \label{eq:pi_plus_theta}
d^{j}_{mk}(\pi+\theta) = (-1)^{j+m} d^{j}_{-m,k}(\theta),
\end{equation}
\begin{equation} \label{eq:pi_minus_theta}
d^{j}_{mk}(\pi-\theta) = (-1)^{j+m} d^{j}_{m,-k}(\theta).
\end{equation}



%

\end{document}